\documentclass[sigconf]{acmart}

\usepackage{multicol}
\usepackage{multirow}
\usepackage{listings}

\definecolor{findingsbg}{RGB}{245,245,245}

\AtBeginDocument{%
  }

\copyrightyear{2026}
\acmYear{2026}
\setcopyright{cc}
\setcctype{by-nc-nd}
\acmConference[MSSiS '26]{8th Workshop on Modeling and Simulation of Software-Intensive Systems }{April 12--18, 2026}{Rio de Janeiro, Brazil}
\acmBooktitle{8th Workshop on Modeling and Simulation of Software-Intensive Systems (MSSiS '26), April 12--18, 2026, Rio de Janeiro, Brazil}

\acmDOI{10.1145/3786146.3788643}
\acmISBN{979-8-4007-2380-3/2026/04}

\begin{document}

\title{Class Model Generation from Requirements using Large Language Models}

\author{Jackson Nguyen}
\authornote{First two authors contributed equally to this research.}
\email{jngu0076@student.monash.edu}
\affiliation{%
  \institution{Monash University}
  \city{Melbourne}
  \state{VIC}
  \country{Australia}
}

\author{Rui En Koe}
\email{rkoe0003@student.monash.edu}
\affiliation{%
  \institution{Monash University}
  \city{Melbourne}
  \state{VIC}
  \country{Australia}
}

\author{Fanyu Wang}
\email{fanyu.wang@monash.edu}
\affiliation{%
  \institution{Monash University}
  \city{Melbourne}
  \state{VIC}
  \country{Australia}
}

\author{Chetan Arora}
\email{chetan.arora@monash.edu}
\affiliation{%
  \institution{Monash University}
  \city{Melbourne}
  \state{VIC}
  \country{Australia}
}

\author{Alessio Ferrari}
\email{alessio.ferrari@ucd.ie}
\affiliation{%
  \institution{University College Dublin}
  \city{Dublin}
  \country{Ireland}
}

\renewcommand{\shortauthors}{Nguyen, Koe, et al.}

\begin{abstract}
The emergence of Large Language Models (LLMs) has opened new opportunities to automate software engineering activities that traditionally require substantial manual effort. Among these, class diagram generation represents a critical yet resource-intensive phase in software design. This paper investigates the capabilities of state-of-the-art LLMs, including GPT-5, Claude Sonnet 4.0, Gemini 2.5 Flash Thinking, and Llama-3.1-8B-Instruct, to generate UML class diagrams from natural language requirements automatically. To evaluate the effectiveness and reliability of LLM-based model generation, we propose a comprehensive dual-validation framework that integrates an \textit{LLM-as-a-Judge} methodology with human-in-the-loop assessment. Using eight heterogeneous datasets, we apply chain-of-thought prompting to extract domain entities, attributes, and associations, generating corresponding PlantUML representations. The resulting models are evaluated across five quality dimensions: completeness, correctness, conformance to standards, comprehensibility, and terminological alignment. Two independent LLM judges (Grok and Mistral) perform structured pairwise comparisons, and their judgments are further validated against expert evaluations. Our results demonstrate that LLMs can generate structurally coherent and semantically meaningful UML diagrams, achieving substantial alignment with human evaluators. The consistency observed between LLM-based and human-based assessments highlights the potential of LLMs not only as modeling assistants but also as reliable evaluators in automated requirements engineering workflows, offering practical insights into the capabilities and limitations of LLM-driven UML class diagram automation.

\end{abstract}



\maketitle

\section{Introduction}
\label{sec:introduction}

Graphical models are a crucial step in {Requirements Engineering} (RE) by helping stakeholders visualize, communicate, and verify system requirements \cite{ferrari2024model}. 
Among many modeling languages, the Unified Modeling Language (UML) \cite{larman1998applying} has become one of the most widely adopted for representing structural and behavioral aspects of systems. {However, designing detailed UML diagrams generally requires significant manual effort and expertise. Moreover, the elicitation of diagrams requires a comprehensive understanding of domain knowledge. These issues can lead to misunderstandings between requirements engineers and stakeholders, further complicating the UML modelling process \cite{tororequirements}.}


Generative AI (GenAI) and Large Language Models (LLMs) have emerged as powerful tools for automating software development, particularly for tasks such as requirements analysis, model generation, and traceability. LLMs can bridge informal requirements with formal models and diagrams by processing natural language inputs and generating human-like responses. This capability could significantly reduce the tedious task of manual modeling and make RE more accessible to non-technical stakeholders. Despite the potential of LLMs, generating UML models from natural language (NL) requirements remains under-explored. Most previous research has focused on areas such as requirements summarization, traceability, and extraction. Earlier attempts at automating model generation relied on rule-based natural language processing (NLP) techniques, which often lacked the flexibility to generalize component interactions for more specialized domains \cite{meng2024automated}.

The application of LLMs to automated UML generation raises two fundamental questions that are critical for software engineers seeking to adopt these technologies. First, \textit{when we use LLMs to generate UML models, can they perform well?} This question directly addresses the practical utility of LLMs for RE practitioners who need reliable, accurate diagrams to support system design and stakeholder communication. Second, \textit{when we use LLMs to evaluate UML models, can they understand them well?} This question becomes essential when considering scalable evaluation approaches, particularly given the challenge of validating generated diagrams without ground truth reference models. These two questions are intrinsically connected and equally important for establishing trust in LLM-based automation. Software engineers need confidence not only that LLMs can generate quality diagrams, but also that automated evaluation methods can reliably assess that quality. Understanding both capabilities is essential for determining whether LLMs can be deployed in real-world RE workflows.

In this paper, we investigate the capabilities of state-of-the-art LLMs (ChatGPT's GPT-5, Claude Sonnet 4.0, Gemini 2.5 Flash Thinking, and Llama-3.1-8B-Instruct) in generating, understanding, and evaluating UML class diagrams from NL requirements. Given the absence of ground truth in real-world scenarios, we adopt a comprehensive dual-validation approach that combines the LLM-as-a-judge methodology with human expert assessment. Recent research has demonstrated the effectiveness of LLMs as evaluators through pairwise comparisons and ranking methods \cite{dhurandhar-etal-2024-ranking}, \cite{liusie2023llm}, providing consistent evaluation when reference models are unavailable. We extend this approach by validating LLM-based assessments against human expert judgment, establishing the reliability and trustworthiness of automated evaluation for UML diagram generation. Our investigation addresses following RQs\footnote{Artefacts available at \url{https://github.com/jackson0076/FIT4701-GenAI/}}:
\begin{itemize}
    \item \textbf{RQ1: How effectively can LLMs generate UML class diagrams, and can LLMs reliably distinguish quality differences through pairwise evaluation?} To explore the generation ability of LLMs, we extract domain entities, attributes, and associations, and generate PlantUML diagrams for eight heterogeneous datasets using LLMs. To investigate understanding ability, we use additional LLMs to perform pairwise evaluations to determine which LLM produced the most accurate diagrams. 
    \item \textbf{RQ2: To what extent can LLMs help in UML generation and evaluation from human experts' opinions?} Building on the generated output from RQ1, we assign a human-in-the-loop evaluation to validate further the performance of the selected best LLM from RQ1. Furthermore, we quantitatively analyze the alignment between LLMs and human experts to provide in-depth insights into overall LLM capabilities in UML diagrams.
\end{itemize}

%
%

\noindent\textit{\textbf{Structure.}} Section \ref{sec:design} presents the research design and methodology. Section \ref{sec:results} presents the results of RQs 1 and 2. Section \ref{sec:threats} reviews the threats to validity and limitations.

\section{Research Design}
\label{sec:design}

This research aims to understand and evaluate the effectiveness of LLMs in model generation. While LLMs have demonstrated strong performance across a wide range of general-purpose tasks, this study aims to conduct a deeper analysis of how specific models differ in quality and performance, and how effectively they can handle diverse requirements. This work builds on previous work~\cite{ferrari2024model}, which examined the effectiveness of a single model for processing requirements.
This study evaluates the quality of UML class diagrams generated by LLMs from NL requirements. We considered GPT-5, Claude Sonnet 4, Gemini 2.5 Flash Thinking, and Llama-3.1-8B-Instruct\footnote{GPT-5: \url{https://chat.openai.com/}; Claude: \url{https://claude.ai/}; Gemini: \url{https://gemini.google.com/app}; Llama: \url{https://www.llama.com/llama-downloads/}} for accessibility and usability. 

\subsection{\textbf{Datasets}} The datasets were selected from real-world NL requirements across different domains. These datasets were selected as they are well-suited for model generation tasks and reflect industry-relevant requirements commonly encountered in requirements engineering. We selected documents from (i) the `Ten Lockheed Martin Cyber-Physical Challenges'\footnote{\url{https://github.com/hbourbouh/lm_challenges}} with 10 requirements documents from cyber-physical domain; (ii) A dataset of user stories~\cite{dalpiaz2020conceptualizing}, and (iii) the Pure Dataset~\cite{ferrari2017pure}.
The criteria for selecting these documents are (i) \textit{Diversity in domain and structure:} The requirements cover a wide variety of domains,
as well as different types of requirements, focusing on ``shall'' type and user stories; and
(ii) \textit{Consistency with prior work:} The selected requirements were previously used in an established study~\cite{ferrari2024model} and were retained in their original form.

\begin{table}[t]
\footnotesize
\caption{Data Collection Results}
\vspace*{-1em}
\label{tab:data-col}
\centering
\begin{tabular}{@{}lll@{}}
\toprule
\textbf{File}$^{*}$ & \textbf{Domain} & \textbf{REQ}$^{\dag}$ \\ 
\midrule
g14-datahub (us) & Data Management & 67 \\
g04-recycling (us) & Recycling System & 51 \\
g12-camperplus (us) & Camping System & 13 \\
UHOPE (us) & Healthcare & 12 \\
Autopilot (s) & Cyber-physical System & 14 \\
Finite State Machine (s) & Embedded Systems & 13 \\
Inventory (s) & Inventory System & 22 \\
Pacemaker (s) & Medical Devices & 187 \\
\bottomrule
\end{tabular}

\footnotesize
$^{\dag}$ REQ: the number of analysed requirements.\\
$^{*}$ us: user stories; s: 'shall' requirements.
\vspace*{-1.5em}
\end{table}

\subsection{Statistical Analysis Approach}

We employ multiple complementary statistical measures to assess agreement and consistency throughout our evaluation:

\noindent$\bullet$ \textbf{Spearman rank correlation ($\rho$)}: Measures monotonic agreement in model rankings (ordinal consistency). Used in RQ1 to assess whether two judges consistently rank models in the same order (e.g., both rank GPT-5 first, Claude second). Values close to 1 indicate strong positive correlation; values close to 0 indicate no correlation.

\noindent$\bullet$ \textbf{Cohen's Kappa ($\kappa$)}: Measures categorical agreement beyond chance. We use binary classification (scores 1-3 as ``unacceptable" and 4-5 as ``acceptable") to assess whether evaluators agree on diagram acceptability. This approach is used in both RQ1 (inter-judge agreement) and RQ2 (LLM-human alignment).

\noindent$\bullet$ \textbf{Statistical significance tests}: Before computing effect sizes, we test for statistical significance of score differences. We first assess normality using Shapiro-Wilk test ($\alpha = 0.05$). For normally distributed data, we apply paired t-tests; for non-normal distributions, we use Wilcoxon signed-rank tests. This ensures appropriate statistical inference based on data characteristics.

\noindent$\bullet$ \textbf{Cohen's d}: Quantifies the magnitude (effect size) of score differences between evaluators, regardless of statistical significance. This provides practical significance beyond p-values, indicating whether differences are meaningful in practice. Thresholds: small (d = 0.2), medium (d = 0.5), large (d = 0.8), very large (d $\geq$ 1.0).

This multi-faceted approach provides comprehensive assessment: Spearman correlation answers "Do judges agree on relative ordering?", Cohen's Kappa answers "Do judges agree on acceptability classification?", statistical tests answer "Are score differences statistically significant?", and Cohen's d answers "How large are the differences in practical terms?" While the specific combination of measures varies based on the research question, the core measurement approaches (categorical agreement via Kappa and effect size via Cohen's d) remain consistent across both RQs.

\subsection{\textbf{RQ1 Design}}
\label{sec:rq1}

\textbf{RQ1} systematically investigates how effectively LLMs can generate and understand UML class diagrams from NL requirements.

\vspace*{0.2em}\noindent\textbf{UML Diagram Generation}. We employ prompting techniques across four LLMs to generate class diagrams across requirements datasets. For each dataset, all LLMs generate a UML class diagram in PlantUML text, adhering to a strict structure that includes packages, methods, classes, interfaces, etc. The choice of PlantUML\footnote{\url{https://www.planttext.com}/} is based on the previous study~\cite{ferrari2024model} and on its widespread use and ease of interpretation.
To standardise the process, we use the \textit{chain-of-thought prompting pattern}~\cite{zhang2023automatic} to ensure a step-by-step approach for extracting relevant domain entities, attributes, and associations from NL requirements. 

\vspace*{0.2em}\noindent\textbf{Prompt Design}.
{The prompts used in this study were designed through an iterative process based on preliminary experiments and prior work on LLM-based model generation {\cite{dhurandhar-etal-2024-ranking}, \cite{ferrari2024model}, \cite{liusie2023llm}}. Early tests that relied on open-ended instructions or \textit{visualization generator prompting pattern} would often produce incomplete diagrams and syntactically incorrect PlantUML outputs. This process would encourage the introduction of more strict structural guidelines and constraints in the prompt, where the prompts were designed to perform the modelling task into explicit steps.
Relative ranking is prioritised in RQ1 to address calibration challenges in the absence of ground truth, as comparative judgements are more robust for identifying consistent performance differences among models.
RQ2 focuses on absolute scoring to validate alignment between the LLM and human evaluations. To reduce scoring noise and improve consistency, scores are assigned on ordinal 1–5 scale rather than a fine-grained numerical range, making judgments easier to calibrate and compare across evaluators. }
The prompt:
\begin{lstlisting}
You are a senior software architect.
 Let's think step-by-step from the requirements to:
 (a) extract entities, roles, packages; (b) define attributes (with types) and methods (signatures/returns) of each entity; (c) decide inheritance, interfaces, enums; (d) assign associations with multiplicities + labels; (e) sanity-check PlantUML syntax so it would compile.
 Then output only the final PlantUML code, wrapped inside one single fenced code block
Requirements: {paste requirements here}
Constraints:
Organise classes into packages (add others if useful)..
Use inheritance where appropriate (e.g., Publisher extends User).
All attributes and methods must include a concrete type.
Visibility (+/#/-) should be included where meaningful; default is + (public) if not specified in the requirements or examples.
Relationships show multiplicities (e.g., "1", "0..*", etc) and labels where meaningful.
Avoid placeholders like <Type>--use concrete names.
Keep design-specific infra (DB tables, REST endpoints) out of the class model.
Do not include any text after @enduml. An example PlantUML structure to follow:
@startuml
package "gui" {
	Game --> "1" Player_manager
	Game --> "1" Display_manager
    
	abstract Updatable

	Players_layout --|> Updatable
	Player_manager --|> Updatable
	Display_manager --|> Updatable
	Character --|> Updatable
	InputBox --|> Updatable

	abstract Gui_control

	Players_layout --|> Gui_control
	Player_manager --|> Gui_control
	Display_manager --> "1" InputBox
	Display_manager --> "1" Player_manager 
	Display_manager --> "1" Players_layout
	Player_manager --> "1..*" Character
}

package "connection" {
	Network --> "1" ConnectionManager
}

package "model" {
	Game_state --> "1..*" Player
	Game_state --> "*" Collectable
}	

class Updatable {
	void update()
}
@enduml
Output: one PlantUML code block only (from `@startuml` to `@enduml`)
\end{lstlisting}

Evaluation reliability depends on clear prompting and well-defined criteria, which were provided to guide the LLM in assigning scores and ranking diagrams in a structured manner. After performing the pairwise comparisons and obtaining the rankings for the dataset, a Spearman correlation analysis is conducted to evaluate the degree of association between the two judge LLMs. To assess the correlation between Grok and Mistral, we apply the Spearman rank correlation coefficient, defined as:
\begin{align}
    \rho = 1 - \frac{6 \sum d_i^2}{n(n^2 - 1)}
\end{align}

where $d_i$ is the difference between the ranks assigned by Grok and Mistral for item $i$, $n$ is the total number of ranked items, $\sum d_i^2$ is the sum of squared rank differences across all the different items.

\noindent The Spearman correlation coefficient ($\rho$) measures the degree of monotonic relationship between two variables, which in this case are the two rankings. A value closer to 1 implies a strong positive correlation, whereas -1 implies a strong negative correlation.

\vspace*{0.2em}\noindent\textbf{LLM-as-a-Judge Evaluation}. To evaluate the understanding capabilities of LLMs, our study employs a reproducible \textit{LLM-as-a-judge} approach, where external LLMs are prompted to perform structured pairwise comparisons of the generated diagrams. We apply a five-criterion evaluation framework, adapted from prior work~\cite{ferrari2024model}, to assess the quality of generated UML diagrams. The criteria are \textbf{Completeness:} The diagram covers the aspects of the requirements with a sufficient degree of detail to communicate with potential stakeholders. \textbf{Correctness:} The diagram specifies a behaviour that accurately and logically reflects the requirements. \textbf{Adherence to standards:} The diagram is syntactically and semantically correct (i.e., it can be interpreted by PlantText). \textbf{Comprehensibility:} The diagram is sufficiently clear given the complexities of the requirements, and is understandable from a stakeholder perspective. \textbf{Terminological alignment:} The terminology used in the diagram and code aligns closely with that used in the requirements. 

{In rare cases, it is possible for pairwise comparisons to result in a tie. Because the web-based LLM interfaces used in this study do not expose confidence scores, confidence-based tie breaking was not possible. In those cases, the tied outputs were re-evaluated by human evaluators using the same evaluation criteria, and the evaluators compared the tied outputs across all criteria and selected the superior model based on overall alignment, ensuring a fair resolution of ties without inconsistent decision rules.
}

For our study, we selected Grok and Mistral Small 3.1 24B as evaluation judges. These models were chosen to provide autonomous judgment and to reduce bias from relying on a single judge. Their role is to compare the class diagrams generated by four candidate LLMs and determine which model produced the most accurate, well-structured diagram, based on the five criteria. We decided to include a locally hosted LLM (Mistral) as one of the judges because we believe local models can enhance reproducibility and accessibility, allowing evaluations to be replicated without relying on external infrastructure. Furthermore, the two models were chosen to reduce bias by ensuring that evaluations were independent of the model families used for generation. This diversity in evaluators strengthens the reliability while ensuring that our method remains transparent and adaptable for future studies. The prompt:

\begin{lstlisting}
Role: You are a strict evaluator of UML class diagrams written in PlantUML. Compare two candidate diagrams (A and B) against the given Context.
Context: <requirements> 
PlantUML result A: <Code for PlantUML A> 
PlantUML result B: <Code for PlantUML B> 

Evaluation rules:
- Consider only UML class diagram semantics (eg. classes, attributes, associations, multiplicities, generalisation, composition/aggregation). 
- Ignore layout, styling, skinparams, comments, and notes. 
- Treat the Context as authoritative; do not invent entities or relationships not present there.

Decision criteria (in priority order):
1. Correct domain classes and relationships
2. Correct associations/multiplicities
3. Appropriate attributes and types
4. Terminology alignment
5. UML class diagram conformance

Evaluation Scores: Assign a score from 1-5 for each criterion when comparing and evaluating the two models. The scoring must strictly adhere to the metrics below, while following the evaluation rules and decision criteria.

Evaluation Metrics:
1. Appropriate conceptual classes aligned with the domain
1 - Very Poor: Classes are mostly incorrect or irrelevant and do not accurately represent the domain.
2 - Poor: Only a few classes are correct, but the majority of the domain concepts are missing or incorrect. 
3 - Fair: Some classes are correctly identified, but noticeable misalignments are still present.
4 - Good: Most classes are correctly identified with only minor misalignments
5 - Excellent: All classes are correctly identified, fully aligned with the domain. 

2. Associations/Multiplicities between classes for domain reasoning 
1 - Very Poor: Associations/multiplicities are largely missing or incorrect, comprising domain reasoning. 
2 - Poor: Few correct associations/multiplicities with critical relationships missing. 
3 - Fair: Some associations/multiplicities are correct, but noticeable missing or misrepresentations remain. 
4 - Good: Most associations/multiplicities are correctly identified with only minor errors. 
5 - Excellent: All essential associations/multiplicities correctly applied.

3. Model attributes with appropriate data types
1 - Very Poor: Attributes and data types are poorly identified and modeled; largely incorrect or irrelevant.
2 -  Poor: Few attributes are correct with proper types; critical attributes are missing.
3 - Fair: Some attributes are correct with proper types, but significant attributes are still missing.
4 - Good: Most attributes are appropriately modeled with proper data types; minor type issues may exist.
5 - Excellent: All attributes are correctly modeled with appropriate data types.

4. Consistencies of Terminology with given context
1 - Very Poor: Terminology is largely inconsistent with the given context 
2 - Poor: Frequent inconsistencies; critical terms misused or missing. 
3 - Fair: Some terminology is correct, but noticeable inconsistencies remain.
4 - Good: Most terminology aligns with the given context, with minor inconsistencies. 
5 - Excellent: Terminology is fully consistent with the given context. 

5. Adherence of UML class diagram standards 
1 - Very Poor: Poor diagram formation with major UML standards ignored, making it largely incomprehensible.
2 - Poor: Frequent UML violations; diagram is difficult to interpret.
3 - Fair: Some UML/PlantUML syntax errors, but major representations are still missing. 
4 - Good: Minor UML/PlantUML notation errors; overall diagram remains understandable.
5 - Excellent: Fully adheres to UML/PlantUML syntax, notation, and conventions.

Return only: Winner: A | B, and Justification: 2-4 concise sentences citing concrete elements from A and B, and includes the scores assigned to the criteria above
\end{lstlisting}

To further assess the degree of consistency between the two LLM judges (\textbf{Grok} and \textbf{Mistral}), we compute effect sizes using Cohen's \textit{d}~\cite{cohen2013statistical}. Cohen's \textit{d} quantifies the magnitude of the difference between their evaluation scores, providing insight into how similarly both judges assessed the generated UML diagrams. Smaller effect sizes indicate stronger alignment (i.e., closer agreement between Grok and Mistral), whereas larger values reflect a greater divergence in the scoring behaviour. The following thresholds are commonly accepted to interpret Cohen's \textit{d} values:

$\bullet$ \textbf{0.2} -- Small effect: Minor difference between means.

$\bullet$ \textbf{0.5} -- Medium effect: Moderate difference between means.

$\bullet$ \textbf{0.8} -- Large effect: Substantial difference between means.

$\bullet$ \textbf{1.0 or greater} -- Very large effect: Major divergence.

\subsection{\textbf{RQ2 Design}}

Building on the generated outputs from RQ1, we select the best-performing LLM identified through the pairwise evaluations to conduct a comprehensive human-in-the-loop (HITL) validation. This research question investigates the extent to which LLMs can assist in generating and evaluating UML from human experts' opinions by conducting expert evaluations and quantitatively analysing the alignment between LLM judges and human evaluators.

\vspace*{0.2em}\noindent\textbf{Human Expert Evaluation}. To validate the performance of the selected best-performing LLM, {we conducted a human-in-the-loop evaluation using an absolute scoring rubric aligned with the quality dimensions assessed in RQ1}. Two independent human evaluators (A1 and A2) were recruited based on their expertise in software engineering and requirements modeling. Both evaluators possess advanced degrees in computer science and have professional experience in UML modeling and requirements analysis, ensuring they have the necessary domain knowledge to perform reliable assessments. Prior to the formal evaluation, both evaluators underwent an education session where they were briefed on the scoring rubric to ensure consistent interpretation of the five criteria: completeness, correctness, adherence to standards, comprehensibility, and terminological alignment. During this session, the evaluators reviewed sample diagrams and discussed the scoring guidelines to establish a shared understanding of what constitutes each score level (1-5) for each criterion. This calibration process was designed to minimize subjective bias and enhance inter-rater reliability. Following the calibration, each evaluator independently assessed the UML diagrams generated by the best-performing model for all eight datasets. The evaluators were provided with the original NL requirements alongside the corresponding PlantUML diagrams. They were instructed to assign scores from 1 (Very Poor) to 5 (Excellent) for each of the five criteria, following the detailed rubric descriptions. {Human evaluators were presented with visual UML diagrams rather than raw PlantUML text, reflecting how UML models are typically interpreted in practical software design environments.} To maintain evaluation independence and prevent anchor bias, evaluators worked separately without cross-discussion during the assessment process. Each evaluator documented their scores and provided brief justifications for their ratings, particularly when assigning lower scores or identifying specific issues in the diagrams. 
{The human evaluation process was designed to mirror the structure of the LLM judge evaluation, enabling direct comparison between human and LLM assessments. However compared to RQ1, this focuses on an absolute scoring rubric to analyse quality of results.}
While the study~\cite{ferrari2024model} used manual human assessment, recent work~\cite{dhurandhar-etal-2024-ranking} demonstrates that using LLMs as judges is an effective alternative approach, provided that the scoring framework is well defined. 

\vspace*{0.2em}\noindent\textbf{LLM Re-evaluation and Alignment Analysis}. In parallel with the human evaluation, the LLM judges (Grok and Mistral) re-evaluate the same diagrams from the best-performing model using the same five-criterion rubric. The prompt:

\begin{lstlisting}
Context: <requirements>
Result: <PlantUML>

You are a senior software architect. Can you grade a PlantUML class diagram code against a NL specification context? Assign a score using integers 1-5 for the following criterias, given their rationale. Be strict.

1) Coverage of Requirements (Completeness)
1 - Very Poor: <40% of critical items present; major gaps.
2 - Poor: ~40-59% present; many critical omissions.
3 - Fair: ~60-79% present; several important omissions.
4 - Good: ~80-94% present; a few moderate omissions.
5 - Excellent: >=95% of critical items included; only minor/low-priority gaps.

2) Accuracy & Logical Consistency
1 - Very Poor: Misrepresents domain; incoherent or contradictory.
2 - Poor: Multiple major mismatches (roles, multiplicities, directions).
3 - Fair: Several inaccuracies but overall intent mostly intact.
4 - Good: One or two minor slips; generally faithful and consistent.
5 - Excellent: Fully consistent with Context; correct types, cardinalities, inheritance, and relation kinds.

3) UML & PlantUML Standards (syntax + semantics)
1 - Very Poor: Broadly invalid UML/PlantUML.
2 - Poor: Likely won't compile or meaning is distorted.
3 - Fair: Probably compiles; multiple style/semantic issues (missing types/visibilities, messy relations).
4 - Good: Compiles; only minor style nits.
5 - Excellent: Clean, compilable; correct packages/classes/enums/interfaces, visibilities, types, labeled relations, proper inheritance/interfaces.

4) Understandability (stakeholder clarity)
1 - Very Poor: diagram cluttered, confusing or ambiguous (names/relations/responsibilities); hard to follow/understand/interpret
2 - Poor: Some diagram's elements are understandable, but overall clarity is low due to poor organisations or missing elements
3 - Fair: Mixed clarity; readable with effort.
4 - Good: Mostly clear and readable (naming, structure, and labels) with logical layout and minimal ambiguity. 
5 - Excellent: Very clear, well-organised packages (e.g., UserManagement, DataPackageManagement, Registry), helpful labels, minimal clutter.

5) Terminology Alignment (matches requirement wording)
1 - Very Poor: <25% alignment; frequent misleading renames.
2 - Poor: ~25-49% alignment; hard to map terms.
3 - Fair: ~50-74% alignment; several renames/synonyms.
4 - Good: ~75-89% alignment; minor, traceable renames.
5 - Excellent: >=90% key terms preserved or transparently mapped.

\end{lstlisting}

This human-in-the-loop component serves as the baseline for quantifying agreement between the LLM judges and human evaluations. To quantify the alignment between LLM judges and human evaluators, we apply Cohen's Kappa ($\kappa$) to measure agreement and Cohen's \textit{d} to estimate the effect size of score differences between human and LLM assessments. To measure inter-rater reliability, the raw ratings from Grok, Mistral, and human evaluators (A1,A2) were binarised into ``unacceptable" and "acceptable" scores. For each group (Grok and Mistral), and for evaluators A1 and A2, an overall consensus label was calculated using an OR rule: an item was marked acceptable if either evaluator had deemed it acceptable, and unacceptable only when both had rated it low. This conservative strategy avoids penalising edge disagreements in judgements while identifying the clear negative consensus for each score, thereby computing a Cohen's Kappa ($\kappa$) score. This approach enables a direct comparison of how closely LLM-based evaluations approximate human-level judgements in UML diagram review tasks, providing in-depth insights into LLMs' overall capabilities for generating and evaluating UML diagrams.


\section{Execution and Results}
\label{sec:results}


\subsection{RQ1 Analysis} 
\begin{figure}
     \centering
\includegraphics[width=0.95\linewidth]{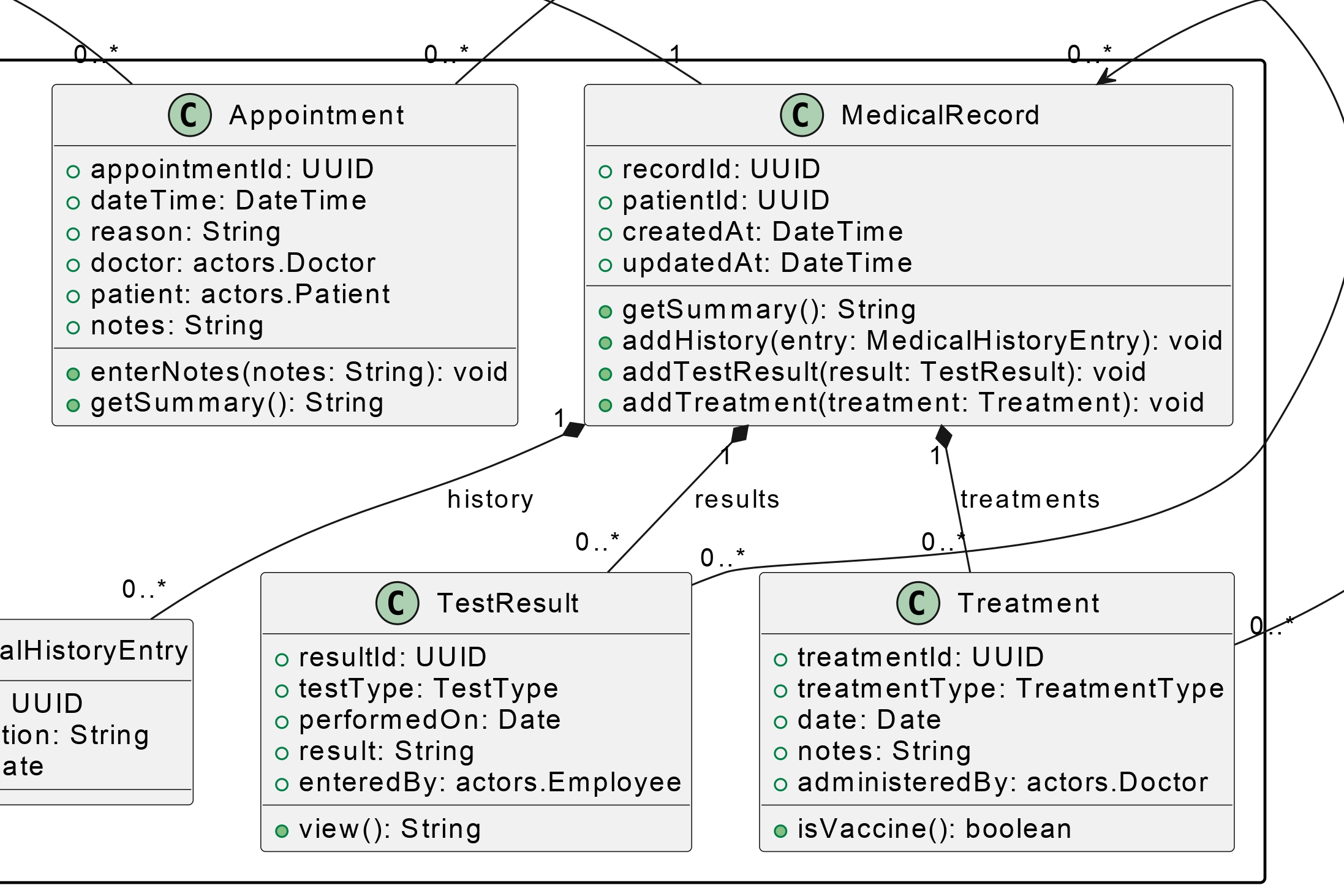}
        \vspace*{-1em}
        \caption{Examples of generated UML class diagrams (GPT-5).}
        \label{fig:model-generation}
        \vspace*{-1em}
\end{figure}

RQ1 systematically investigates LLMs' performance across generation and evaluation. In a practical scenario, these two capabilities are significant for software developers: the generation enables developers to create class diagrams without reading lengthy NL requirements, and the evaluation enables both technical and non-technical users to assess diagram quality without expert supervision.
\noindent\textbf{UML Diagram Generation} We generated class diagrams using four evaluated LLMs with chain-of-thought prompting across eight datasets (see example in Fig.~\ref{fig:model-generation}). Overall, these LLMs demonstrate their ability to effectively extract domain concepts and generate corresponding model diagrams from NL requirements.

\noindent\textbf{Error Analysis} As illustrated in Fig.~\ref{fig:model-generation}, LLM-generated diagrams generally capture core domain classes but occasionally have missing or incorrect associations and multiplicities, redundant associations.  The major structural errors were consistently identified by the LLM judges and human evaluators, whereas minor and more subtle issues were less penalised. This observation explains the variability in the understandability criteria and highlights possible areas for improvement in diagram generation and evaluation.


\noindent\textbf{LLM-as-a-Judge Pairwise Evaluation}. To validate the capability of LLMs in evaluating the best-performing model in the absence of reference diagrams, we selected two independent LLM judges (Grok and Mistral) to perform pairwise comparisons using the five criteria: completeness, correctness, adherence to standards, comprehensibility, and terminological alignment. Tables \ref{tab:llm-rankings-grok-mistral}~(1)~and~(2) show the rankings assigned by the two judges across eight datasets, where each LLM diagram was ranked from best (1) to worst (4).

\begin{table}[!t]
\footnotesize
\caption{LLM Rankings judged by Grok and Mistral (1/2)}
\vspace*{-1em}
\label{tab:llm-rankings-grok-mistral}
\centering

\begin{tabular}{|l|cc|cc|cc|cc|}
\hline
\multirow{2}{*}{\textbf{Dataset}} & \multicolumn{2}{c|}{\textbf{GPT}}        & \multicolumn{2}{c|}{\textbf{Claude}}         & \multicolumn{2}{c|}{\textbf{Gemini}}         & \multicolumn{2}{c|}{\textbf{Llama}}          \\ \cline{2-9} 
                                  & \multicolumn{1}{c|}{\textbf{1}} & \textbf{2} & \multicolumn{1}{c|}{\textbf{1}} & \textbf{2} & \multicolumn{1}{l|}{\textbf{1}} & \textbf{2} & \multicolumn{1}{l|}{\textbf{1}} & \textbf{2} \\ \hline
g14-datahub                       & \multicolumn{1}{c|}{1}          & 1          & \multicolumn{1}{c|}{2}          & 3          & \multicolumn{1}{l|}{3}          & 2          & \multicolumn{1}{l|}{4}          & 4          \\
g04-recycling                     & \multicolumn{1}{c|}{2}          & 2          & \multicolumn{1}{c|}{1}          & 1          & \multicolumn{1}{l|}{3}          & 3          & \multicolumn{1}{l|}{4}          & 4          \\
g12-camperplus                    & \multicolumn{1}{c|}{3}          & 3          & \multicolumn{1}{c|}{1}          & 1          & \multicolumn{1}{l|}{2}          & 2          & \multicolumn{1}{l|}{4}          & 4          \\
UHOPE                             & \multicolumn{1}{c|}{2}          & 1          & \multicolumn{1}{c|}{1}          & 2          & \multicolumn{1}{l|}{3}          & 3          & \multicolumn{1}{l|}{4}          & 4          \\
autopilot                         & \multicolumn{1}{c|}{1}          & 1          & \multicolumn{1}{c|}{2}          & 2          & \multicolumn{1}{l|}{3}          & 3          & \multicolumn{1}{l|}{4}          & 4          \\
FiniteStateMachine                & \multicolumn{1}{c|}{1}          & 2          & \multicolumn{1}{c|}{2}          & 1          & \multicolumn{1}{l|}{3}          & 3          & \multicolumn{1}{l|}{4}          & 4          \\
Inventory                         & \multicolumn{1}{c|}{1}          & 1          & \multicolumn{1}{c|}{2}          & 2          & \multicolumn{1}{l|}{3}          & 3          & \multicolumn{1}{l|}{4}          & 4          \\
Pacemaker                         & \multicolumn{1}{c|}{1}          & 3          & \multicolumn{1}{c|}{2}          & 1          & \multicolumn{1}{l|}{3}          & 2          & \multicolumn{1}{l|}{4}          & 4          \\ \hline
\end{tabular}
\vspace*{-1em}
\end{table}

Across both judges, \textbf{ChatGPT} consistently ranked highest, indicating strong performance in generating structurally correct and complete UML diagrams. \textbf{Claude} ranked second competitively, while \textbf{Gemini} trailed behind. \textbf{Llama} consistently ranked lowest across both judges' evaluations.

{\noindent\textbf{Inter-Judge Consistency Analysis}}. Table~\ref{tab:spearman} reports the Spearman rank correlation coefficients ($\rho$) between Grok and Mistral. The results indicate strong positive correlations across most datasets, with $\rho$ values ranging from 0.8 to 1.0 for seven out of eight datasets, suggesting high agreement between the two judges. However, the Pacemaker dataset shows a considerably lower correlation ($\rho = 0.2$), suggesting greater variance in rankings. This discrepancy may stem from the dataset's unique domain complexity or differences in how each judge interprets diagram semantics. 
Table~\ref{tab:fivecriteria-merged} show the detailed scores for the best-performing LLM (GPT-5) across all datasets. GPT-5 consistently achieved high scores (4 to 5) for completeness, correctness, and terminology alignment across most datasets, with lower scores observed in Pacemaker and g12-camperplus due to domain complexity and understandability challenges. To assess consistency between the two LLM judges, we computed Cohen's Kappa \cite{kohen1960coefficient} using binary classification (scores 1 to 3 as unacceptable, 4 to 5 as acceptable). The resulting $\kappa = 0.773$ indicates {substantial agreement}, reinforcing the reliability of the LLM-as-a-judge framework.
Table \ref{tab:stats-merged-comparison} shows that effect sizes (Cohen's d) between the two judges were predominantly small, indicating consistent application of the rubric. The notable exception was understandability (d = 0.86), suggesting that readability and clarity can be interpreted differently among LLM judges. For terminology alignment, both judges produced consistent scores with a one-point difference (Grok = 5, Mistral = 4), resulting in an undefined Cohen's d but indicating a consistent scoring offset. {Overall, these results suggest that LLM judges are highly consistent when evaluating concrete structural aspects of UML diagrams, while higher-level interpretive criteria introduce greater variability.
}

{\noindent\textbf{Statistical Significance Testing}}.
We also applied the non-parametric Wilcoxon signed-rank test for each evaluation criterion to verify whether the average scores assigned by the LLM judges differed significantly from the neutral mean value of 3, as done in \cite{ferrari2024model}, \cite{abrahao2011evaluating}. 
The null hypothesis stated that the median score for each criterion does not differ from 3, considering $\alpha = 0.05$. 
The test results indicated that the majority of the criteria were statistically higher than the midpoint, suggesting a high degree of fulfilment in completeness, correctness, adherence to standards, understandability and terminological alignment. 
These outcomes reinforce the consistency observed through Cohen’s Kappa ($\kappa = 0.773$) and the small to medium effect sizes reported earlier, confirming that both LLM judges systematically evaluated the GPT-5 diagrams above the acceptable quality threshold.

\vspace{0.3em}\noindent\fcolorbox{black}{findingsbg}{%
  \begin{minipage}{0.95\linewidth}%
    \vspace{0.1em}\footnotesize\textbf{Answers to RQ1:} LLMs can effectively both generate and evaluate UML class diagrams from NL requirements. GPT-5 consistently outperformed other models across diverse domains. Critically, the LLM-as-a-judge approach demonstrates that two independent judges (Grok and Mistral) can reliably distinguish performance differences among generator models, achieving substantial categorical agreement ($\kappa = 0.773$) and strong ranking correlations across 7 of 8 datasets. Statistical significance testing confirms that both judges systematically evaluated diagrams above the acceptable quality threshold with predominantly small effect sizes, indicating consistent rubric application. However, domain-specific complexity and subjective criteria (understandability: $d = 0.86$) present challenges.\vspace{0.1em}%
  \end{minipage}%
}

\begin{table}[!t]
\small
\caption{Evaluation of \textbf{GPT-5}. \textbf{Grok/Mistral} (1/2) and \textbf{Evaluator A1/A2} (3/4)}
\label{tab:fivecriteria-merged}
\centering
\vspace*{-1em}
\setlength{\tabcolsep}{4pt}
\resizebox{\columnwidth}{!}{%

\begin{tabular}{|l|cc|cc|cc|cc|cc|cc|}
\hline
\multirow{2}{*}{\textbf{Dataset}} & \multicolumn{2}{c|}{\textbf{C1}} & \multicolumn{2}{c|}{\textbf{C2}} & \multicolumn{2}{c|}{\textbf{C3}} & \multicolumn{2}{c|}{\textbf{C4}} & \multicolumn{2}{c|}{\textbf{C5}} & \multicolumn{2}{c|}{\textbf{Total}} \\ \cline{2-13} 
 & \textbf{1/2} & \textbf{3/4} & \textbf{1/2} & \textbf{3/4} & \textbf{1/2} & \textbf{3/4} & \textbf{1/2} & \textbf{3/4} & \textbf{1/2} & \textbf{3/4} & \textbf{1/2} & \textbf{3/4} \\ \hline
g14-datahub & 5/4 & 4/5 & 4/4 & 4/4 & 5/4 & 4/4 & 5/4 & 4/5 & 5/4 & 5/5 & 24/20 & 21/23 \\
g04-recycling & 5/4 & 4/4 & 5/4 & 4/4 & 5/5 & 5/5 & 5/4 & 4/4 & 5/4 & 5/5 & 25/21 & 22/22 \\
g12-camperplus & 4/4 & 4/4 & 3/4 & 4/4 & 3/4 & 4/4 & 3/3 & 4/4 & 5/4 & 5/5 & 18/19 & 21/21 \\
UHOPE & 5/5 & 4/4 & 5/5 & 4/5 & 5/4 & 4/5 & 5/4 & 5/5 & 5/4 & 5/5 & 25/22 & 22/24 \\
autopilot & 5/4 & 3/4 & 5/4 & 4/4 & 5/4 & 4/4 & 5/4 & 5/4 & 5/4 & 5/4 & 25/20 & 21/20 \\
FiniteState. & 4/4 & 4/4 & 4/4 & 4/4 & 5/5 & 5/5 & 4/4 & 5/3 & 5/4 & 4/4 & 22/21 & 22/20 \\
Inventory & 4/4 & 4/4 & 4/4 & 4/4 & 5/5 & 5/4 & 4/4 & 4/4 & 5/4 & 5/4 & 22/21 & 22/20 \\
Pacemaker & 2/3 & 3/3 & 3/3 & 3/3 & 3/3 & 3/3 & 4/4 & 3/3 & 5/4 & 4/3 & 17/17 & 16/15 \\ \hline
\end{tabular}
}
\footnotesize
C1 = Completeness, C2 = Correctness, C3 = Adherence to Standards, C4 = Comprehensibility, C5 = Terminological alignment
\vspace*{-1.5em}
\end{table}

\begin{table}[ht]
\small
\caption{Spearman correlation between Grok/Mistral}
\label{tab:spearman}
\vspace*{-1em}
\centering
\setlength{\tabcolsep}{2pt}
\resizebox{\columnwidth}{!}{%
\begin{tabular}{|ccccccccc|}
\hline
\multicolumn{1}{|l}{{Dataset}}                                      & \multicolumn{1}{l}{g14-datahub} & \multicolumn{1}{l}{g04-recycl.} & \multicolumn{1}{l}{g12-camper.} & \multicolumn{1}{l}{UHOPE} & \multicolumn{1}{l}{autopilot} & \multicolumn{1}{l}{FiniteState.} & \multicolumn{1}{l}{Inventory} & \multicolumn{1}{l|}{Pacemaker} \\ \hline
{Cor. ($\rho$)} & 0.8                             & 1.0                             & 1.0                             & 0.8                       & 1.0                           & 0.8                              & 1.0                           & 0.2                            \\
{p-value}                                                           & 0.2                             & 0.0                             & 0.0                             & 0.2                       & 0.0                           & 0.2                              & 0.0                           & 0.8                            \\ \hline
\end{tabular}%
}
\vspace{-1em}
\end{table}

\subsection{RQ2 Analysis}


{\noindent\textbf{Human Expert Evaluation}}. Two independent human evaluators (A1 and A2) assessed the eight UML diagrams generated by GPT-5 using the same five-criterion rubric after undergoing a calibration session to ensure consistent interpretation.
Table~\ref{tab:fivecriteria-merged} show the evaluations performed by A1 and A2. Both evaluators assigned consistently high scores, particularly for {adherence to standards} and {terminological alignment}, indicating that the generated diagrams were generally well-structured and coherent. To measure inter-rater agreement, the Cohen's Kappa, $\kappa = 0.684$, indicates \emph{substantial agreement} between evaluators. 

{
The left side in Table~\ref{tab:stats-merged-comparison} shows small to medium effect sizes across most criteria ($d \leq 0.5$), indicating substantial consistency between human evaluators in their assessment of diagram quality, which aligns with trends observed for LLM judges, particularly for criteria such as correctness and adherence to standards. Terminological alignment exhibits a slightly larger effect size ($d = 0.61$), suggesting that experts may apply different thresholds when mapping requirement terminology to model elements, reflecting a higher degree of subjectivity for this criteria.}

{\noindent\textbf{Statistical Significance Testing}}. The Wilcoxon signed-rank test was used to examine whether the ratings of human evaluators for each criterion differed significantly from the neutral midpoint of 3, with $\alpha = 0.05$. 
As in RQ1, results showed that most criteria were significantly above this threshold, indicating that human evaluators consistently perceived the ChatGPT GPT-5's generated UML diagrams as exceeding the ``fair" quality level across all criteria. 
These findings align with substantial inter-rater agreement ($\kappa = 0.684$) and small-to-medium effect sizes observed, validating the reliability of human evaluation and alignment with LLM-based assessments.


\begin{table}[!t]
\footnotesize
\caption{Statistics of LLM Judges \& Human Evaluation}
\vspace{-1em}
\label{tab:stats-merged-comparison}
\centering
\setlength{\tabcolsep}{2pt}
\begin{tabular}{|l|ccc|ccc|}
\hline
\multirow{2}{*}{\textbf{Criterion}} & \multicolumn{3}{c|}{\textbf{LLM Judges}} & \multicolumn{3}{c|}{\textbf{Human Evaluation}} \\ \cline{2-7} 
 & \multicolumn{1}{c|}{\textbf{Mean}} & \multicolumn{1}{c|}{\textbf{Cohen's d}} & \textbf{Eff. S.} & \multicolumn{1}{c|}{\textbf{Mean}} & \multicolumn{1}{c|}{\textbf{Cohen's d}} & \textbf{Eff. S.} \\ \hline
Completeness & \multicolumn{1}{c|}{4.125} & \multicolumn{1}{c|}{0.30} & small & \multicolumn{1}{c|}{3.875} & \multicolumn{1}{c|}{-0.50} & medium \\
Correctness & \multicolumn{1}{c|}{4.062} & \multicolumn{1}{c|}{0.18} & none & \multicolumn{1}{c|}{3.938} & \multicolumn{1}{c|}{-0.28} & small \\
Standards & \multicolumn{1}{c|}{4.375} & \multicolumn{1}{c|}{0.30} & small & \multicolumn{1}{c|}{4.250} & \multicolumn{1}{c|}{0.00} & none \\
Understanding & \multicolumn{1}{c|}{4.125} & \multicolumn{1}{c|}{0.86} & large & \multicolumn{1}{c|}{4.125} & \multicolumn{1}{c|}{0.34} & small \\
Terminology & \multicolumn{1}{c|}{4.500} & \multicolumn{1}{c|}{inf} & large & \multicolumn{1}{c|}{4.562} & \multicolumn{1}{c|}{0.61} & medium \\ \hline
\end{tabular}
\vspace*{-1em}
\end{table}

\begin{figure}
    \centering
\includegraphics[width=0.95\linewidth]{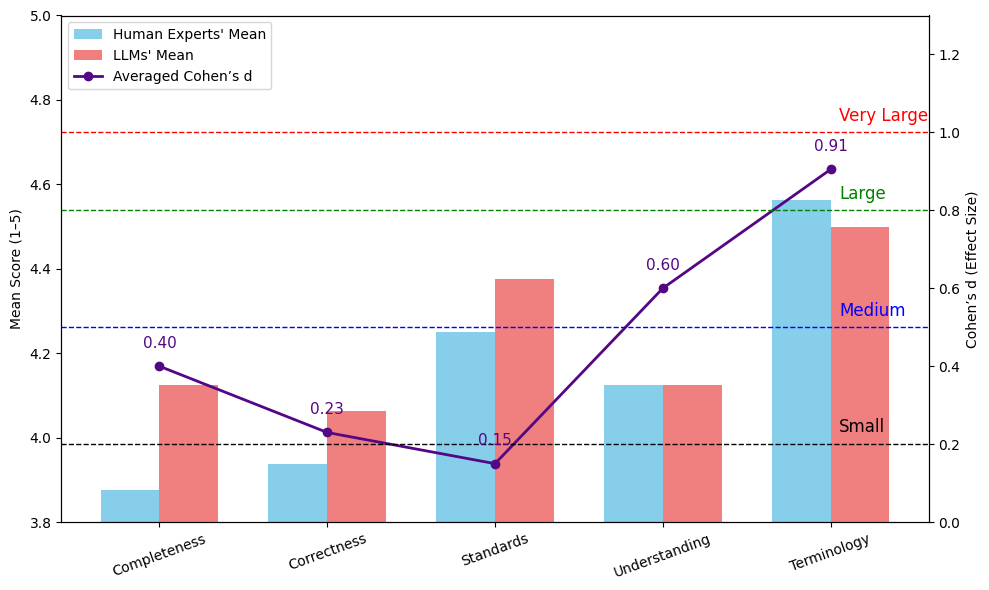}
\vspace*{-1em}
    \caption{Agreement between LLMs and human evaluators.}
    \label{fig:llm-human-alignment}
    \vspace*{-1em}
\end{figure}

{\noindent\textbf{Alignment Analysis}}. Fig.~\ref{fig:llm-human-alignment} presents combination of a bar chart and line plot showing agreement patterns across the five criteria for both LLM judges and human evaluators, where we use mean values for the bar chart, while the overall approximate Cohen’s d is calculated by averaging the Cohen’s d between LLM judges and human evaluators using data from Tables \ref{tab:stats-merged-comparison} for each criterion. Observing Fig. \ref{fig:llm-human-alignment}., the bars are generally close together across all criteria, indicating a high alignment of scoring between LLMs and human evaluators. Both groups assigned the highest ratings to terminology alignment and understandability. Specifically, for understandability, the bars are identical, suggesting strong agreement between LLMs and humans on this criterion. LLM judges tended to assign slightly higher scores than human evaluators for completeness, correctness, and adherence to standards, resulting in small differences in judgment. These minor variations indicate no major bias or deviation between LLM and human assessments. The Kappa values (LLM judges: $\kappa = 0.773$; human evaluators: $\kappa = 0.684$) further substantiate high alignment between the two evaluation approaches. Comparing the overall approximate Cohen’s d for each criterion shown in Fig. \ref{fig:llm-human-alignment}., most criteria exhibit strong alignment, except for understandability and terminological alignment, which show larger effect sizes, suggesting potential differences in interpretation between the groups. 
Finally, another Cohen's $\kappa$ was computed for the inter-rater reliability between the LLM judges and human evaluators. Comparing the aggregated LLM consensus formed using an OR rule, the resulting score yielded $\kappa$ = 0.7222, indicating \emph{substantial agreement between automated and human judgements}.

Overall, the results indicate that LLM judgments closely align with human evaluations, with minor deviations arising from subjective interpretation of certain criteria.  
These findings demonstrate that LLMs can serve as effective preliminary assessors for UML quality evaluation, reducing cognitive and time burdens on human reviewers. The substantial agreement levels validate the reliability of the evaluation framework. The tendency for LLMs to assign higher scores indicates that final validation should remain human-guided, particularly for domain-intensive models. This human-AI collaborative framework offers a promising direction, combining LLM efficiency with human expertise. 

\vspace{0.3em}\noindent\fcolorbox{black}{findingsbg}{%
  \begin{minipage}{0.95\linewidth}%
    \vspace{0.1em}\footnotesize\textbf{Answers to RQ2:} LLMs can significantly assist in UML generation and evaluation from human experts' perspectives. Human evaluators validated that GPT-5 produces high-quality diagrams with substantial inter-rater agreement ($\kappa = 0.684$) and scores significantly above acceptable thresholds (Wilcoxon $p < 0.05$). Strong alignment between LLM judges and human evaluators is evidenced by comparable categorical agreement levels (LLM-LLM: $\kappa = 0.773$; Human-Human: $\kappa = 0.684$), similar mean scores across all criteria, and predominantly small effect sizes. While LLM judges assigned slightly higher scores for completeness, correctness, and adherence to standards, both groups converged on the highest ratings for terminology alignment and understandability. Larger effect sizes for understandability and terminology ($d > 0.6$) reflect differences in subjective interpretation.\vspace{0.1em}%
  \end{minipage}%
}

\section{Threats to Validity}
\label{sec:threats}

\textit{\textbf{Internal Validity.}} LLM judges may rate diagrams differently depending on how they interpret the prompts. Although clear scoring rules were defined, each model (Grok and Mistral) could have its own bias or understanding of UML syntax, potentially affecting how fairly they compare the results. To mitigate this threat, we selected two judge models from separate families. We validated their consistency using various correlation tests (Spearman's correlation and Cohen's Kappa), which demonstrated substantial agreement ($\rho$ ranging from 0.8 to 1.0 across seven of eight datasets; $\kappa = 0.773$). \textit{\textbf{Construct Validity.}} The five-criterion evaluation framework provides a solid overview but may not cover all intricacies that make a model practical and reliable in real projects, such as efficiency or reusability. Furthermore, both human evaluators (A1 and A2) and LLM judges may interpret the scoring criteria slightly differently even with clearly defined criteria, which may affect result reliability. This can be due to training bias or differences in understanding for each model. To address this threat, we measured inter-rater agreement using Cohen's Kappa for both LLM judges ($\kappa = 0.773$) and human evaluators ($\kappa = 0.684$), which showed substantial agreement in both cases. \textit{\textbf{External Validity.}} This study focuses on specific LLM versions. The results across other LLMs or future versions of these models may vary, as training datasets and hyperparameters may be tuned differently. To reduce potential bias from model behaviour, we selected a wide variety of datasets from different domains, ensuring the evaluation was not overly focused on a single domain and that models were tested on a broader range of requirement types. Future work should expand this by testing different requirement types, additional datasets, and more LLMs to validate whether the observed trends remain consistent over time.
\section{Related Work}
\label{sec:related-work}

{The application of LLMs in requirements engineering has gained significant traction, driven by their ability to capture semantic knowledge and observe regularities in natural language \cite{min2023recent}. However, the quality of the input requirements remains a critical bottleneck. Ferrari et al. \cite{ferrari2024model} demonstrated that ambiguities and inconsistencies in requirements frequently lead to omitted elements and structural defects in generated diagrams. Similarly, Vogelsang et al. \cite{vogelsang2024specifications} compared performance across 94 paired requirements, finding that ``requirement smells'', particularly semantic inconsistencies, severely hamper binary tracing and model fidelity. While these studies collectively establish that high-quality inputs are a prerequisite for effective model generation, they primarily focus on the LLM as a \textit{generator}. They do not address the complementary potential of LLMs to act as \textit{judges} of these artifacts, nor do they evaluate how automated quality assessments align with human expertise in the specific context of UML class diagrams.}

{To address the challenges of assessing generative outputs without fixed ground truth, recent research has turned to reference-free evaluation methods. Dhurandhar et al. \cite{dhurandhar-etal-2024-ranking} introduced a triplet-based method where LLMs rank peer models to identify weaker performers, while Liusie et al. \cite{liusie2023llm} validated the use of pairwise comparative assessments for general response quality. Although these approaches demonstrate the feasibility of "LLM-as-a-judge" frameworks, they have largely been tested on general natural language tasks. There remains a significant gap in applying these techniques to software modeling, where correctness is defined by strict structural and terminological standards rather than linguistic fluency. Furthermore, these existing frameworks lack rigorous validation against human expert judgments in the software engineering domain, leaving an open question regarding their reliability for technical artifacts. Our work extends the state of the art by strictly adapting these evaluation techniques to the domain of UML class diagrams. Unlike previous studies, we jointly examine model generation and evaluation, providing a critical analysis of the alignment between LLM-based judges and human assessments to determine the viability of automated evaluation in software design.}

\section{Conclusion}
\label{sec:conclusion}

This study investigated the ability of LLMs to automatically generate and evaluate UML class diagrams from natural language requirements without ground truth data. We compared four LLMs (GPT-5, Claude Sonnet 4.0, Gemini 2.5 Flash Thinking, and Llama-3.1-8B-Instruct) across eight diverse datasets using a five-criterion evaluation framework. Employing the LLM-as-a-Judge approach with two independent evaluators, we found that LLM judges can reliably distinguish performance differences among generator models with substantial agreement ($\kappa = 0.773$) and strong ranking correlations ($\rho = 0.8$ to 1.0 across seven datasets). Human-in-the-loop validation further demonstrated strong alignment between LLM-based and human evaluations ($\kappa = 0.684$ for human evaluators), indicating that LLM judges can approximate human-level judgment when guided by structured rubrics. GPT-5 consistently produced the most accurate diagrams, though domain-specific complexity (e.g., Pacemaker, g12-camperplus) remains challenging for both generation and evaluation.

This work contributes a reproducible framework for assessing LLM-generated UML models and demonstrates the potential of LLM-based tools for both technical and non-technical stakeholders. The findings validate a practical human-AI collaborative approach where LLMs handle initial diagram generation and assessment, while human expertise guides final validation for complex or domain-intensive models.
{From a simulation-oriented perspective, this work supports early stage system analysis rather than simulation execution itself. By automating the generation and evaluation of UML class diagrams from natural language requirements, the framework can assist in producing validated structural models that commonly introduce simulation-based analysis.}

Future work will incorporate additional LLMs, expand dataset diversity, and explore adaptive techniques such as retrieval-augmented generation to improve accuracy and reliability. This framework can be extended to other modeling languages beyond UML diagrams.

\bibliographystyle{ACM-Reference-Format}

\bibliography{mybib}

\end{document}